\documentclass[letterpaper,twocolumn,10pt]{article}
\usepackage{usenix2021}

\usepackage{hyperref}
\usepackage{epsfig,endnotes}
\usepackage{amsmath}
\usepackage{amssymb}
\usepackage{algorithmic}
\usepackage[linesnumbered,ruled]{algorithm2e}

\usepackage{mathtools}
\usepackage{graphicx}
\usepackage[capitalize]{cleveref}
\usepackage{xspace}
\usepackage[small,bf,skip=2pt]{caption}
\usepackage{textcomp}
\usepackage{comment}
\usepackage{booktabs}
\usepackage{subcaption}
\usepackage{fancyhdr}
\usepackage{todonotes}
\usepackage[scaled=0.8]{beramono} 

\usepackage[compact]{titlesec}

\usepackage{enumitem}
\setitemize{noitemsep,topsep=0pt,parsep=0pt,partopsep=0pt}
\setenumerate{noitemsep,topsep=0pt,parsep=0pt,partopsep=0pt}
\setdescription{itemsep=0pt,topsep=0pt,parsep=0pt,partopsep=0pt,itemindent=0em,leftmargin=0.5em}

\newcommand{\us}{\textmu{}s\xspace}

\def\eg{{e.g.\ }}

\usepackage{authblk}
\makeatletter
\renewcommand\AB@affilsepx{, \protect\Affilfont}
\makeatother

\usepackage{tikz}

\usepackage{framed}

\newif\ifxetexorluatex
\ifxetex
  \xetexorluatextrue
\else
  \ifluatex
    \xetexorluatextrue
  \else
    \xetexorluatexfalse
  \fi
\fi
\ifxetexorluatex%
  \usepackage{fontspec}
  \usepackage{libertine} 
  \newfontfamily\quotefont[Ligatures=TeX]{Linux Libertine O} 
\else
  \usepackage[utf8]{inputenc}
  \usepackage[T1]{fontenc}
  \usepackage{libertine} 
  \newcommand*\quotefont{\fontfamily{LinuxLibertineT-LF}} 
\fi

\newcommand*\quotesize{30} 
\newcommand*{\openquote}
   {\tikz[remember picture,overlay,xshift=-4ex,yshift=-2.5ex]
   \node (OQ) {\quotefont\fontsize{\quotesize}{\quotesize}\selectfont``};\kern0pt}

\newcommand*{\closequote}[1]
  {\tikz[remember picture,overlay,xshift=4ex,yshift={#1}]
   \node (CQ) {\quotefont\fontsize{\quotesize}{\quotesize}\selectfont''};}

\colorlet{shadecolor}{white}

\newcommand*\shadedauthorformat{\emph} 

\newcommand*\authoralign[1]{%
  \if#1l
    \def\authorfill{}\def\quotefill{\hfill}
  \else
    \if#1r
      \def\authorfill{\hfill}\def\quotefill{}
    \else
      \if#1c
        \gdef\authorfill{\hfill}\def\quotefill{\hfill}
      \else\typeout{Invalid option}
      \fi
    \fi
  \fi}
\newenvironment{shadequote}[2][l]%
{\authoralign{#1}
\ifblank{#2}
   {\def\shadequoteauthor{}\def\yshift{-2ex}\def\quotefill{\hfill}}
   {\def\shadequoteauthor{\par\authorfill\shadedauthorformat{#2}}\def\yshift{2ex}}
\begin{snugshade}\begin{quote}\openquote}
{\shadequoteauthor\quotefill\closequote{\yshift}\end{quote}\end{snugshade}}

\usepackage{etoolbox}
\makeatletter
\patchcmd{\maketitle}
	{\@maketitle}
	{\@maketitle\vspace{-2em}}
	{}
	{}
\makeatother

\begin{document}

\title{BPF for storage: an exokernel-inspired approach\vspace{-0.6cm}}
\author[1]{Yu Jian Wu}
\author[1]{Hongyi Wang}
\author[1]{Yuhong Zhong}
\author[1]{\authorcr Asaf Cidon}
\author[2]{Ryan Stutsman}
\author[3]{Amy Tai}
\author[1]{Junfeng Yang}
\affil[1]{Columbia University}
\affil[2]{University of Utah}
\affil[3]{VMware Research}

\maketitle

\begin{abstract}

The overhead of the kernel storage path accounts
for half of the access latency for new NVMe storage devices.
We explore using BPF to reduce this overhead, by injecting user-defined functions deep in the kernel's I/O processing stack.
When issuing a series of dependent I/O requests, this approach can increase IOPS by over 2.5$\times$ and cut latency by half, by bypassing
kernel layers and avoiding user-kernel boundary crossings.
However, we must avoid losing important properties when bypassing the file system and block layer such as the safety guarantees of the file system and translation between physical blocks addresses and file offsets.
We sketch potential solutions to these problems, inspired by exokernel file systems from the late 90s, whose time, we believe, has finally come!
\vspace{-0.3cm}
\begin{shadequote}[r]{Attributed to King Solomon}
As a dog returns to his vomit, so a fool repeats his folly.
\vspace{-0.1cm}
\end{shadequote}
\vspace{-0.4cm}

\end{abstract}

\section{Introduction}

Storage devices have historically lagged behind networking devices in
achieving high bandwidth and low latency. While 100~Gb/s bandwidth is now
common for network interface cards (NICs) and the physical layer, storage
devices are only beginning to support 2-7~GB/s bandwidth and 4-5~\us{}
latencies~\cite{z-nand,xl-flash,alderstream-datasheet,xpoint-gen2}. With such
devices, the software stack is now a substantial overhead on every storage
request. In our experiments this can account for about half of I/O operation
latency, and 
the impact on
throughput can be even more significant.

Kernel-bypass frameworks
(e.g.\ SPDK~\cite{yang2017spdk}) and near-storage processing reduce kernel overheads.
However, there are clear drawbacks to both
such as significant, often bespoke, application-level
changes~\cite{ruan2019insider,qiao2018high}, lack of isolation, wasteful busy waiting when I/O usage isn't high,
and the need for specialized hardware in
the case of computational storage~\cite{ngd-storage,smartssd,willow}.
Therefore, we want a standard OS-supported mechanism that can reduce the software overhead for fast storage devices.

To address this, we turn to the networking community, which has long had high-bandwidth devices.
Linux's eBPF~\cite{ebpf} provides an interface for applications to embed simple functions
directly in the kernel.
When used to intercept I/O, 
these functions can perform processing that is traditionally done in the application and can avoid
having to copy data and incurring context switches when going back and forth between the kernel and user space.
Linux eBPF is widely used for packet processing and filtering~\cite{hoiland2018express,cloudflare-ebpf},
security~\cite{MAC-ebpf} and tracing~\cite{gregg2019bpf}.

BPF's ubiquity in the kernel and its wide acceptance make it a natural scheme for
application-specific kernel-side extensions in layers outside of the networking stack.
BPF could be used to chain dependent I/Os, eliminating 
costly traversals of the kernel storage stack and transitions to/from userspace. For example, it could be used to
traverse a disk-based data structure like a B-tree where one block references another.
Embedding these BPF functions deep enough in the kernel has the potential to
eliminate nearly all of the software overhead of issuing I/Os like
kernel-bypass, but, unlike kernel-bypass, it does not require polling and wasted CPU time.

To realize the performance gains of BPF we identify four substantial open research challenges which are unique to the
storage use case. 
First, for ease of adoption, our architecture must support Linux with standard file systems and applications
with minimal modifications to either. It should also be efficient and bypass as many software layers as feasible.
Second, storage pushes BPF beyond current, simple packet
processing uses. Packets are self-describing, so BPF can operate on
them mostly in isolation. Traversing a structure
on-disk is stateful and frequently requires consulting outside state.
Storage BPF functions will need to understand applications'
on-disk formats and access outside state in the
application or kernel, for example, to synchronize concurrent accesses or to
access in-memory structure metadata.
Third, we need to ensure that BPF storage functions cannot violate file system security guarantees while still allowing sharing of disk capacity among applications.
Storage blocks themselves typically do not record ownership or access control attributes, in contrast to
network packets whose headers specify the flows to which the packets belong.
Hence, we need an efficient scheme for enforcing access control that doesn't induce the full cost of the kernel's file system and block layers.
Fourth, we need to enable concurrency.
Applications support concurrent accesses via fine-grained synchronization
(\eg{} lock coupling~\cite{btree-survey}) to avoid read-write interference with
high throughput; synchronization from BPF functions may be needed.


Our approach is inspired by the work on exokernel file system designs.
User-defined kernel extensions were the cornerstone of XN for the Xok
exokernel. It supported mutually distrusting ``libfs''es via code downloaded
from user processes to the kernel~\cite{exokernel}. In XN, these
\emph{untrusted deterministic functions} were interpreted to give the kernel
a user-defined understanding of file system metadata. Then, the application
and kernel design was clean slate, and absolute flexibility in file system
layout was the goal. While we require a similar mechanism to allow users to
enable the kernel to understand their data layout, our goals are different:
we want a design that allows applications to define custom BPF-compatible
data structures and functions for traversing and filtering
on-disk data, works with Linux's existing interfaces and file systems,
and substantially prunes the amount of kernel code executed per-I/O to drive
millions of IOPS.

\section{Software is the Storage Bottleneck}

The past few years have seen new memory technologies emerge in SSDs attached
to high bandwidth PCIe using NVMe. This has led to storage devices that now
rival the performance of fast network devices~\cite{cx6} with a few
microseconds of access latency and gigabytes per second of
bandwidth~\cite{z-nand,xl-flash,alderstream-datasheet,xpoint-gen2}. Hence,
just as the kernel networking stack emerged as a CPU bottleneck for fast
network cards~\cite{erpc,snap,tsai2017lite,IX}, the kernel storage stack is
now becoming a bottleneck for these new devices.

\begin{figure}[t!]
\centering
\includegraphics[width=\columnwidth]{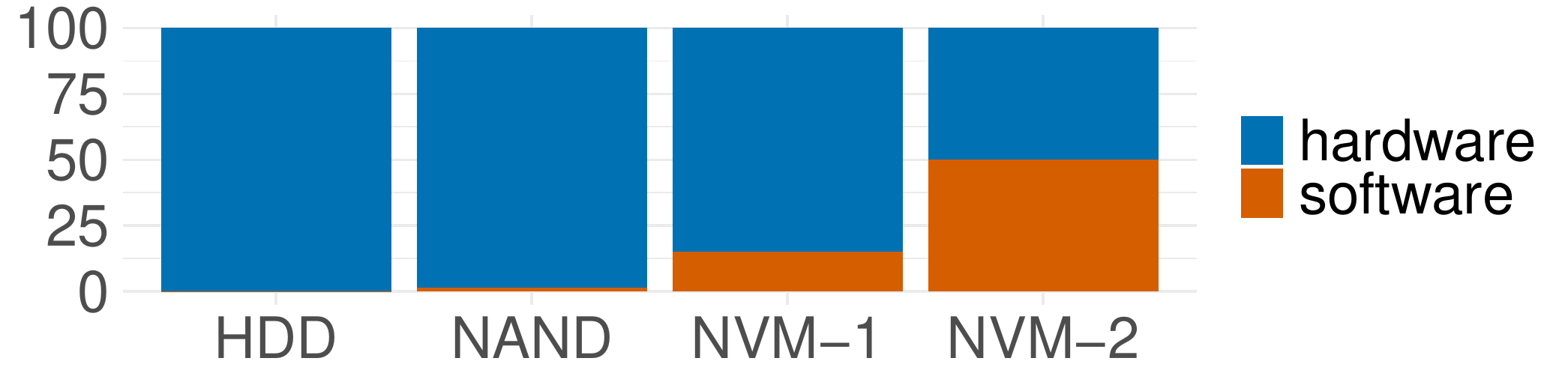}
\caption{Kernel's latency overhead with 512~B random reads. HDD is Seagate Exos X16, NAND is Intel Optane 750 TLC NAND, NVM-1 is first generation Intel Optane SSD (900P), and NVM-2 is second generation Intel Optane SSD (P5800X)}
\label{fig:software_overhead}
 \vspace{-4mm}
\end{figure}

Figure~\ref{fig:software_overhead} shows this; it breaks down the fraction of
read I/O latency that can be attributed to the device hardware and the system
software for increasingly fast storage devices. The results show the kernel's
added software overhead was already measurable (10-15\% latency overhead) with
the first generation of fast NVMe devices (\eg first generation Optane SSD or
Samsung's Z-NAND); \emph{in the new generation of devices software accounts
for about half of the latency of each read I/O}.

\begin{table}[!t]
  \centering
  \begin{tabular}{lrr}
  \toprule
  kernel crossing & 351~ns & 5.6\% \\
  read syscall & 199~ns & 3.2\% \\
  ext4 & 2006~ns & 32.0\% \\
  bio & 379~ns & 6.0\% \\
  NVMe driver & 113~ns & 1.8\% \\
  storage device & 3224~ns & 51.4\% \\
  \midrule
  total & 6.27~\us & 100.0\%\\
  \bottomrule
  \end{tabular}
  \caption{Average latency breakdown of a 512~B random \texttt{read()} syscall using Intel Optane SSD gen 2.}
  \label{tab:latency-breakdown}
\end{table}

\begin{figure}[t]
\centering
\includegraphics[width=\columnwidth]{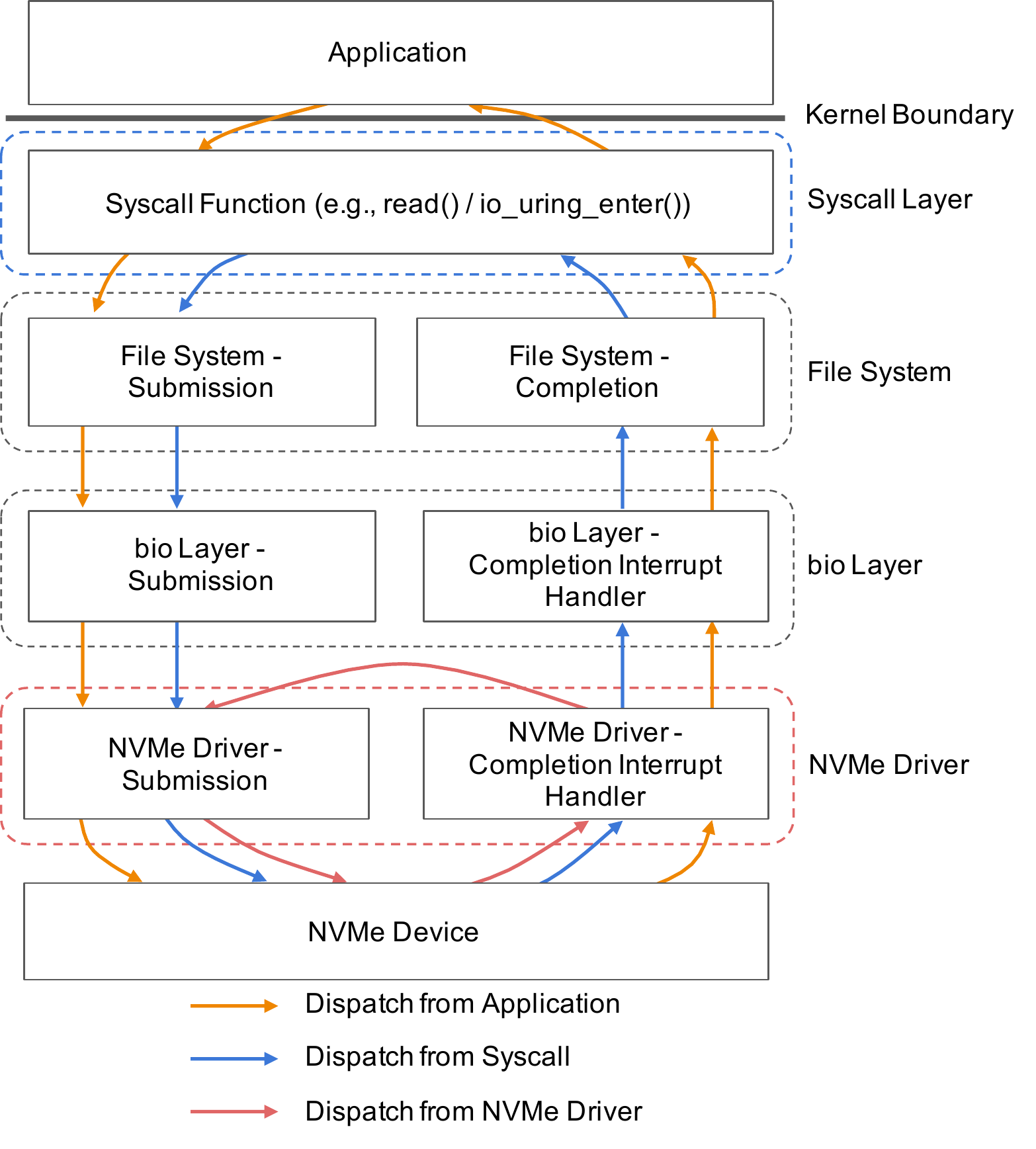}
\caption{Dispatch paths for the application and the two kernel hooks.}
\label{fig:hook}
\vspace{-4mm}
\end{figure}	

\vspace{-1mm}
\paragraph{The Overhead Source.} To breakdown this software overhead, we
measured the average latency of the different software layers when issuing a
random 512~B \texttt{read()} system call with O\_DIRECT on an Intel Optane SSD Gen 2 prototype
(P5800X) on a 6-core i5-8500 3~GHz server with 16~GB of memory, Ubuntu~20.04,
and Linux~5.8.0. We use this setup throughout the paper.
We disable processor C-states and turbo boost and use the
maximum performance governor. Table~\ref{tab:latency-breakdown} shows that
the layers that add the most latency are the file system (ext4 here),
followed by the transition from user space into kernel space.

\begin{figure*}[t]
\centering
\begin{subfigure}[t]{0.24\textwidth}
\includegraphics[width=\columnwidth]{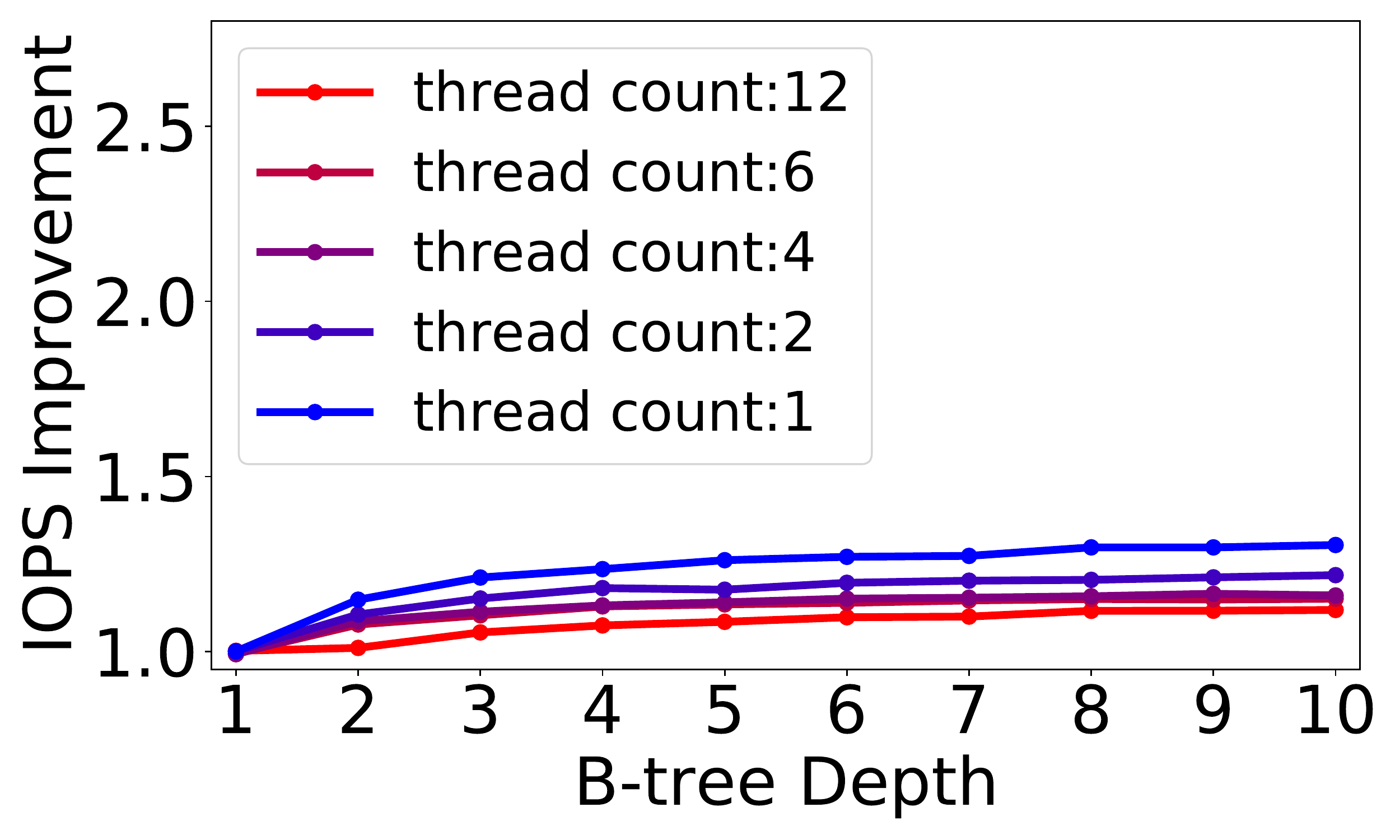}
\caption{Lookups with read syscall, using syscall dispatcher layer hook.}
\label{fig:read_syscall_iops}
\end{subfigure}\hfill
\begin{subfigure}[t]{0.24\textwidth}
\includegraphics[width=\columnwidth]{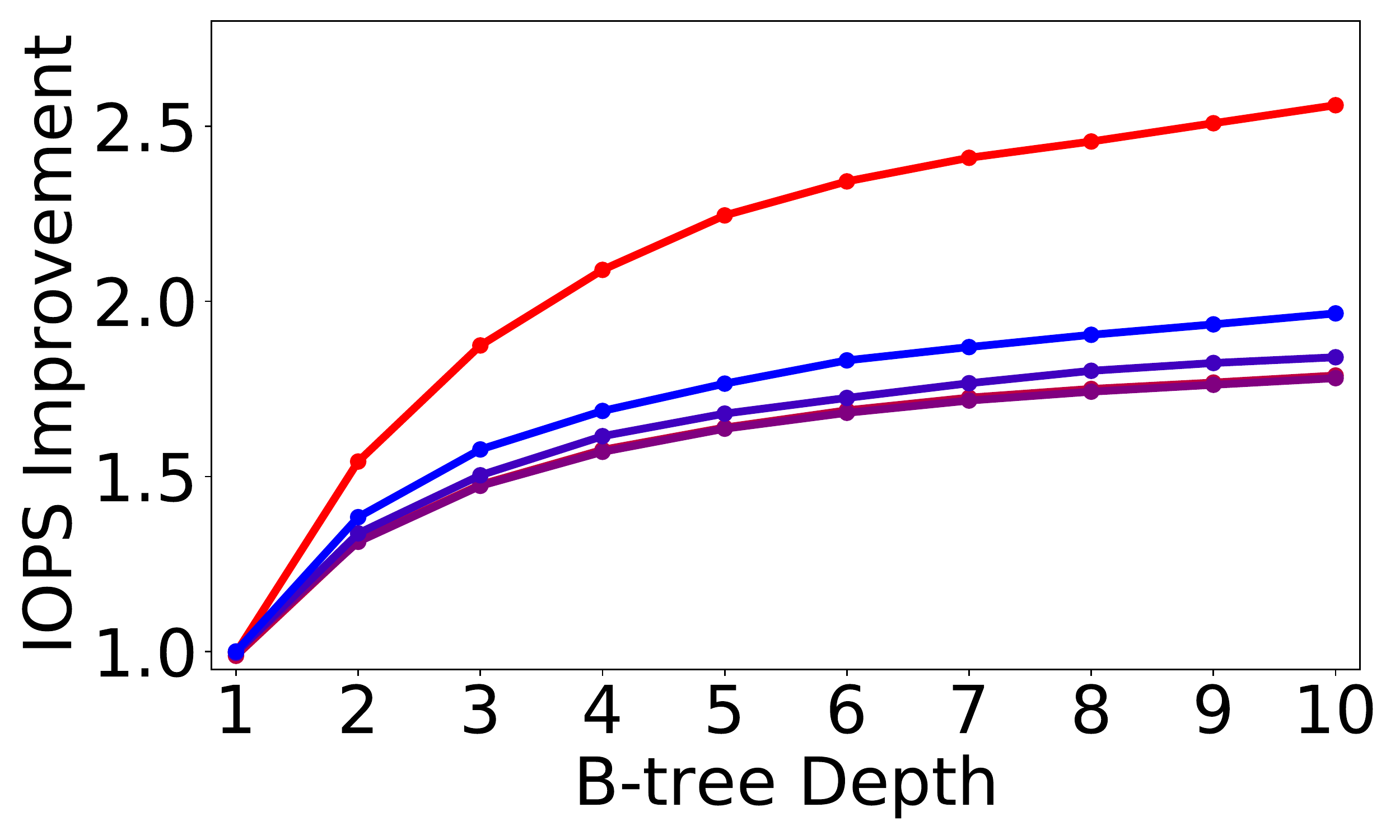}
\caption{Lookups with read syscall, using NVMe driver layer hook.}
\label{fig:read_driver_iops}
\end{subfigure}\hfill
\begin{subfigure}[t]{0.24\textwidth}
\includegraphics[width=\columnwidth]{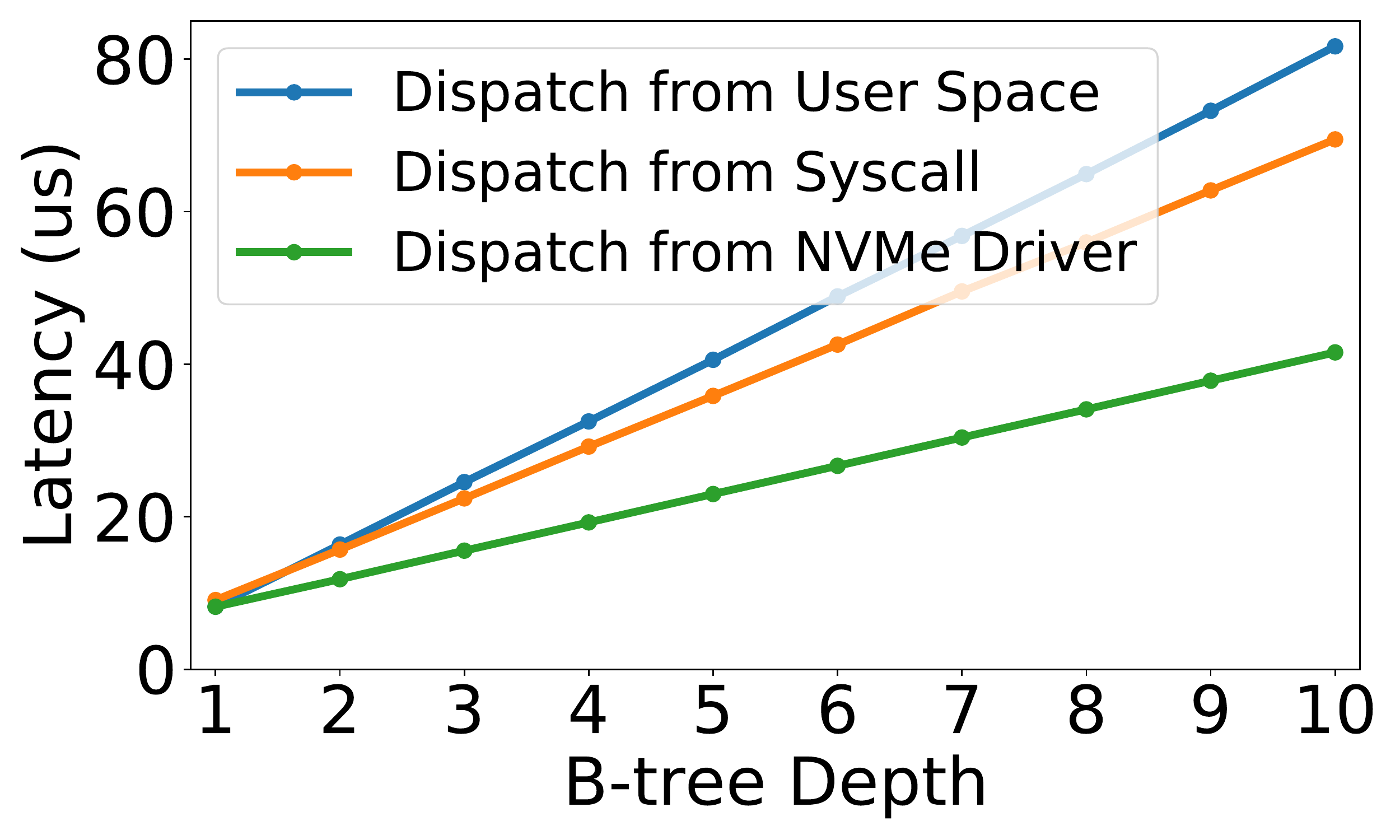}
\caption{One-threaded lookup with read syscall, using syscall layer hook and NVMe driver layer hook.}
\label{fig:read_syscall_latency}
\end{subfigure}\hfill
\begin{subfigure}[t]{0.24\textwidth}
\includegraphics[width=\textwidth]{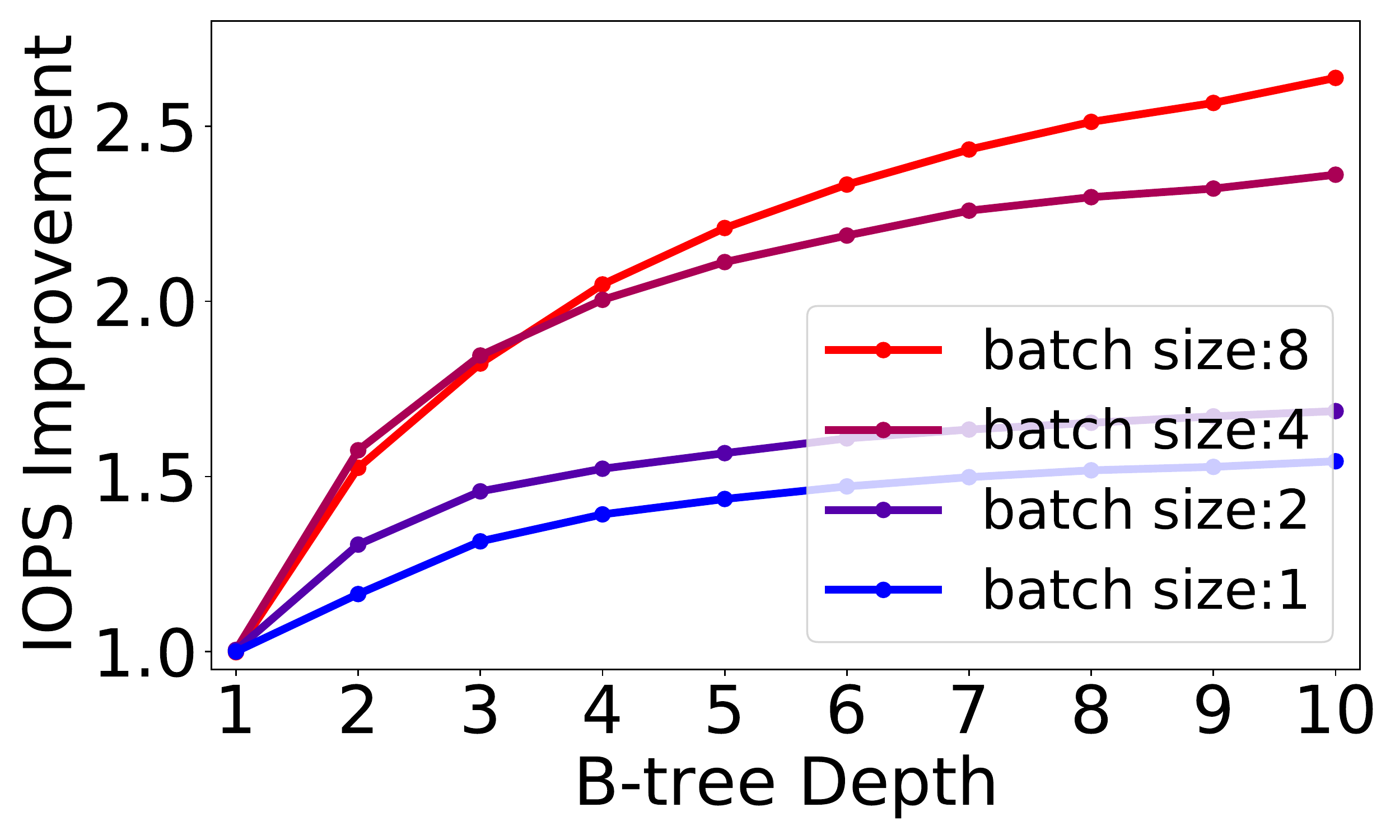}
\caption{Single-threaded lookups with io\_uring syscall, using NVMe driver hook.}
\label{fig:iouring_driver_iops}
\end{subfigure}
\caption{Search throughput improvement on B-tree with varying depth when reissuing lookups from different kernel layers.}
  \vspace{-4mm}
\end{figure*}

Kernel-bypass allows applications directly submit requests to devices,
effectively eliminating all of these costs except the costs to post NVMe
requests (``NVMe driver'') and the device latency
itself~\cite{yang2017spdk,IX,zhang2019,shinjuku}. However, eliminating all of
these layers comes at a high cost. The kernel can only delegate whole devices to
processes, and applications must implement their own file
systems on raw devices~\cite{ruan2019insider,qiao2018high}. Even when they
do, they still cannot safely share files or capacity between distrusting
processes. Finally, lack of efficient application-level interrupt dispatching
means these applications must resort to polling to be efficient at high load;
this makes it impossible for them to efficiently share cores with other
processes, resulting in wasteful busy polling when I/O utilization isn't high
enough. 

There have been recent efforts to streamline kernel I/O submission costs via
a new system call called \emph{io\_uring}~\cite{iouring}. It provides batched
and asynchronous I/O submission/completion path that amortizes the costs of
kernel boundary crossings and can avoid expensive scheduler interactions to
block kernel threads. However, each submitted I/O still must pass through all
of the kernel layers shown in Table~\ref{tab:latency-breakdown}. So, each I/O
still incurs significant software overhead when accessing fast storage
devices (we quantify this in \S\ref{sec:potential}).


\section{BPF to the Rescue?}
\label{sec:potential}

With the rise of fast networks, Berkeley Packet Filter (BPF) has gained in
popularity for efficient packet processing since it eliminates the need to
copy each packet into userspace; instead applications can safely
operate on packets in the kernel. Common networking use
cases include filtering
packets~\cite{xdp,hoiland2018express,cloudflare-ebpf,cilium}, network
tracing~\cite{bpftrace,bcc,gregg2019bpf}, load
balancing~\cite{katran,cilium}, packet steering~\cite{packet-steering} and
network security checks~\cite{MAC-ebpf}. It has also been used as a way to avoid
multiple network crossings when accessing disaggregated storage~\cite{kourtis2020safe}.
Linux supports BPF via the eBPF extension since Linux 3.15~\cite{ebpf-history}.
The user-defined functions can be executed either
using an interpreter or a just-in-time (JIT) compiler.

\vspace{-1mm}
\paragraph{BPF for Storage.}
We envision similar use of BPF for storage by
removing the need to traverse the kernel's storage stack and move data back and forth between the kernel and user
space when issuing dependent storage requests.
Many storage applications consist of many ``auxiliary'' I/O requests, such
as index lookups. A key characteristic of these requests is that
they occupy I/O bandwidth and CPU time to fetch data that is
ultimately \emph{not} returned to the user. For example, a search on
a B-tree index is a series of pointer lookups that lead to the final I/O
request for the user's data page.
Each of these lookups makes a
roundtrip from the application through the kernel's storage
stack, only for the application to throw the data away after simple
processing. Other, similar use cases include database iterators that scan
tables sequentially until an attribute satisfies a
condition~\cite{rocksdb-iterator} or graph databases that execute depth-first
searches~\cite{neorj-dfs}.
%
%

%

\vspace{-1mm}
\paragraph{The Benefits.}
We design a benchmark that executes lookups on an on-disk B$^{+}$-tree (which we call a B-tree for simplicity),
a common data structure used to index databases~\cite{sql-server,idreos2019design}.
For simplicity, our experiments assume that the leaves of the index contain user data rather than
pointers~\cite{lu2017wisckey}.
We assume each node of the B-tree sits on a separate disk page,
which means for a B-tree of depth $d$, a lookup requires reading $d$ pages from disk.
The core operation of a B-tree lookup is parsing the current page to find the offset of the next disk page,
and issuing a read for the next page, a ``pointer lookup''.
Traditionally, a B-tree lookup requires 
$d$ successive pointer lookups from userspace.
To improve on the baseline, we reissue successive pointer lookups
from one of two hooks in the kernel stack: the syscall dispatch layer (which
mainly eliminates kernel boundary crossings)
or the NVMe driver's interrupt handler on the completion path (which
eliminates nearly all of the software layers from the resubmission).
Figure~\ref{fig:hook} shows the dispatch paths for the two hooks along with the normal user space dispatch path.
These two hooks are proxies for the eBPF hooks that we would ultimately use
to offload user-defined functions.

Figures~\ref{fig:read_syscall_iops} and \ref{fig:read_driver_iops} show the throughput speedup
of both hooks
relative to the baseline application traversal, and Figure~\ref{fig:read_syscall_latency} shows the latency
of both hooks while varying the depth of the B-tree.
When lookups are reissued from the syscall dispatch layer, the maximum
speedup is 1.25$\times$.
The improvement is modest because each lookup still incurs the file system and block layer overhead;
the speedup comes exclusively from eliminating kernel boundary crossings.
As storage devices approach 1~\us latencies, we expect greater speedups from this dispatch hook.
On the other hand, reissuing from the NVMe driver makes subsequent I/O requests significantly
less computationally
expensive, by bypassing nearly the entire software stack. 
Doing so achieves speed ups up to 2.5$\times$, and reduces latency by up to 49\%.
Relative throughput improvement actually goes down when adding more threads,
because the baseline application also benefits from more threads
until it reaches CPU saturation at 6 threads.
Once the baseline hits CPU saturation, the computational savings due to 
reissuing at the driver becomes much more apparent.
The throughput improvement from reissuing in the driver continues to scale with deeper trees,
because each level of the tree compounds the number of requests that are issued cheaply.


\paragraph{What about io\_uring?}
The previous experiments use Linux's standard, synchronous
\texttt{read} system call.
Here, we repeat these experiments using the more efficient and
batched io\_uring submission path to drive B-tree lookups from a single thread.
Like before, we reissue lookups from within the NVMe driver and plot the throughput
improvement against an application that simply batches I/Os using unmodified
io\_uring calls. Figure~\ref{fig:iouring_driver_iops} shows the throughput
speedup due to reissuing from within the driver relative to the application
baseline.

As expected, increasing the batch size (number of system calls batched
in each io\_uring call), increases the speedup, since a higher batch size
increases the number of requests that can be reissued
at the driver. For example, for a batch size of 1 only 1 request
(per B-tree level) can be reissued inexpensively, whereas for a batch size of
8, each B-tree level saves on 8 concurrent requests.
%
Therefore, placing the hooks close to the device benefits both standard, synchronous
\texttt{read} calls and more efficient io\_uring calls.
With deep trees, BPF coupled with io\_uring delivers $>$~2.5$\times$
higher throughput; even three dependent lookups give 1.3--1.5$\times$ speedups.


\section{A Design for Storage BPF}

Our experiments have given us reason to be optimistic about BPF's potential
to accelerate operations with fast storage devices; however, to realize these
gains, I/O resubmissions must happen as early as possible, ideally within the
kernel NVMe interrupt handler itself. This creates significant challenges in
using BPF to accelerate storage lookups for a practical system such as a
key-value store.

We envision building a library that provides a higher level-interface than
BPF and new BPF hooks in the Linux kernel as early in the storage I/O
completion path as possible, similar to XDP~\cite{xdp}.
This library would contain BPF functions to accelerate access and operations
on popular data structures, such as B-trees and log-structured merge trees
(LSM).

Within the kernel, these BPF functions that would be triggered in the NVMe
driver interrupt handler on each block I/O completion. By giving these
functions access to raw buffers holding block data, they could extract file
offsets from blocks fetched from storage and immediately reissue an I/O to
those offsets; they could also filter, project, and aggregate block data by
building up buffers that they later return to the application. By pushing
application-defined structures into the kernel these functions can traverse
persistent data structures with limited application involvement. Unlike XN,
where functions were tasked with implementing full systems, these storage BFP
functions would mainly be used to define the layout of a storage blocks that
make up application data structures.

We outline some of the key design considerations and challenges for our
preliminary design, which we believe we can realize without substantial
re-architecture of the Linux kernel.

\vspace{-1mm}
\paragraph{Installation \& Execution.} To accelerate dependent accesses, our
library installs a BPF function using a special \texttt{ioctl}. Once installed, the
application I/Os issued using that file descriptor are ``tagged''; submissions
to the NVMe layer propagate this tag. The kernel I/O completion path, which
is triggered in the NVMe device interrupt handler, checks for this tag. For
each tagged submission/completion, our NVMe interrupt handler hook passes the
read block buffer into the BPF function.

When triggered, the function can perform a few actions. For example, it can
extract a file offset from the block; then, it can ``recycle'' the NVMe
submission descriptor and I/O buffer by calling a helper function that
retargets the descriptor to the new offset and reissues it to the NVMe device
submission queue. Hence, one I/O completion can determine the next I/O that
should be submitted with no extra allocations, CPU cost, or delay. This lets
functions perform rapid traversals of structures without application-level
involvement.

The function can also copy or aggregate data from the block buffer into its
own buffers. This lets the function perform selection, projection, or
aggregation to build results to return to the application. When the function
completes it can indicate which buffer should be returned to the application.
For cases where the function started a new I/O and isn't ready to return
results to the application yet (for example, if it hasn't found the right
block yet), it can return no buffer, preventing the I/O completion from
being raised to the application.

\vspace{-1mm}
\paragraph{Translation \& Security.}
In Linux the NVMe driver doesn't have access to file system metadata. If an
I/O completes for a block at offset $o_1$ in a file, a BPF function
might extract file offset $o_2$ as the next I/O to issue. However, $o_2$ is
meaningless to the NVMe context, since it cannot tell which physical block
this corresponds to without access to the file's metadata and extents. Blocks
could embed physical block addresses to avoid the need to consult the
extents, but without imposing limits on these addresses, BPF functions could
access any block on the device. Hence, a key challenge is imbuing the NVMe
layer with enough information to efficiently and safely map file offsets to
the file's corresponding physical block offsets without restricting the file
system's ability to remap blocks as it chooses.

For simplicity and security in our design, each function only uses the file
offsets in the file to which the \texttt{ioctl} attached the function. This
ensures functions cannot access data that does not belong to the file. To do
this without slow file system layer calls and without constraining the file
system's block allocation policy, we plan to only trigger this block
recycling \emph{when the extents for a file do not change}. We make the
observation that many data center applications do not modify persistent
structures on block storage in place. For example, once an LSM-tree writes
SSTable files to disk, they are immutable and their extents are
stable~\cite{dong2017optimizing}. Similarly, the index file extents remain
nearly stable in on-disk B-tree implementations; In a 24 hour
YCSB~\cite{ycsb} (40\% reads, 40\% updates, 20\% inserts, Zipfian~0.7)
experiment on MariaDB running TokuDB~\cite{tokudb}, we found the index file's
extents only changed every 159~seconds on average with only 5 extent changes
in 24~hours unmapping any blocks. Note that in these index implementations,
each index is stored on a single file, and does not span multiple files, which
further helps simplify our design.

We exploit the relative stability of file extents via a soft state cache of
the extents at the NVMe layer. When the \texttt{ioctl} first installs the
function on the file storing the data structure, its extents are propagated
to the NVMe layer. If any block is unmapped from any of the file's extents, a
new hook in the file system triggers an invalidation call to the NVMe layer.
Ongoing recycled I/Os are then discarded, and an error is returned to
the application layer, which must rerun the \texttt{ioctl} to reset the NVMe
layer extents before it can reissue tagged I/Os. This is a heavy-handed but
simple approach. It leaves the file system almost entirely decoupled from the
NVMe layer, and it places no restrictions on the file system block allocation
policies. Of course, these invalidations need to be rare for the cache to be
effective, but we expect this is true in most of the applications we target.

\vspace{-1mm}
\paragraph{I/O Granularity Mismatches.} When the BIO layer ``splits'' an I/O,
\eg across two discontiguous extents, it
will generate multiple NVMe operations that complete at different times.
We expect these cases to be rare enough that we can perform that I/O as a
normal BIO and return the buffer and completion to the application.
There, it can run the BPF function itself and restart the I/O chain with the
kernel starting at the next ``hop''. This avoids extra
complexity in the kernel. 
Similarly, if application needs to generate more than one I/O in response
to a single I/O completion, we propagate the completion up to
the BIO layer which allocates and submits the multiple I/Os to the NVMe
layer. This avoids returning to userspace.

\vspace{-1mm}
\paragraph{Caching.} As the caching of indices is often managed by the
application~\cite{tokudb,dong2017optimizing,FASTER}, we assume the BPF
traversal will not interact with the buffer cache directly and that
applications manage caching and synchronizing with traversals.
Cache eviction and management is increasingly done at the granularity of
application-meaningful objects (\eg individual data records) instead of whole
pages. Our scheme fits well into this model, where BPF functions can return
specific objects to the application rather than pages, to which it can apply
its own caching policies.

\vspace{-1mm}
\paragraph{Concurrency and Fairness.}
A write issued through the file system might only be reflected in the buffer
cache and would not be visible to the BPF traversal. This could be addressed
by locking, but managing application-level locks from within the NVMe driver
could be expensive. Therefore, data structures that require fine-grained
locking (\eg lock coupling in B+trees~\cite{btree-survey}) require careful
design to support BPF traversal.

To avoid read/write conflicts, we initially plan to target data
structures that remain immutable (at least for a long period of time).
Fortunately, many data structures have this property, including LSM SSTable
files that remain immutable~\cite{o1996log,dong2017optimizing}, and on-disk
B-trees that are not updated dynamically in-place, but rather in a batch
process~\cite{tokudb}. In addition, due to the difficulty of acquiring locks,
we plan initially to only support read-only BPF traversals. 

BPF issued requests do not go through
the file system or block layer, so there is no easy place to enforce fairness or
QoS guarantees among processes. However, the default block layer scheduler in Linux is
the noop scheduler for NVMe devices, and the NVMe specification
supports command arbitration at hardware queues if fairness is a
requirement~\cite{nvme-spec}. Another challenge is that the NVMe layer may
reissue an infinite number of I/O requests. The eBPF verifier prevents loops
with unknown bounds~\cite{bpf-verification}, but we would also need to
prevent unbounded I/O loops at our NVMe hook.

For fairness purposes and to prevent unbounded traversals, we plan to
implement a counter per process in the NVMe layer that will track the number
of chained submissions, and set a bound on this counter. The counter's values
can periodically be passed to the BIO layer to account the number of
requests.

\section{Conclusions}

BPF has the potential to significantly speed up dependent lookups to fast storage devices.
However, it creates several key challenges, arising due to the loss of
context when operating deep in the kernel's storage stack. In this paper, we focused
primarily on enabling the initial use case of index traversals. Notwithstanding, even for current fast NVMe devices (and more so for
future ones), chaining a small
number of requests using BPF provides significant gains. We envision a BPF for storage
library could help developers offload many other standard storage operations to the kernel,
such as compaction, compression, deduplication and scans. We also believe
that the interactions of BPF with the cache and scheduler policies
create exciting future research opportunities.


\section{Acknowledgments}
We would like to thank Frank Hady and Andrew Ruffin for their generous support,
and to Adam Manzanares, Cyril Guyot, Qing Li and Filip Blagojevic for their guidance throughout the project and feedback on the paper.

\bibliographystyle{plain}
\bibliography{database}

\end{document}